\documentstyle[prd,aps,epsf,floats,amsfonts,amssymb,amsmath]{revtex}
\newcommand{\be}{\begin{equation}}\newcommand{\ee}{\end{equation}}
\newcommand{\bea}{\begin{eqnarray} }\newcommand{\eea}{\end{eqnarray}}
\newcommand{\beaa}{\begin{eqnarray}}\newcommand{\eeaa}{\end{eqnarray}}
\newcommand{\ba}{\begin{array}}\newcommand{\ea}{\end{array}}
\newcommand{\bit}{\begin{itemize}}\newcommand{\eit}{\end{itemize}}
\newcommand{\ben}{\begin{enumerate}}\newcommand{\een}{\end{enumerate}}
\newcommand{\bib}{\bibitem}
\def\lab{\label}\def\lan{\langle}
\def\lf{\left}\def\Lrar{\Leftrightarrow}
\def\non{\nonumber}
\def\pa{\partial}\def\ran{\rangle}\def\rar{\rightarrow}
\def\ri{\right}\def\ti{\tilde}

\def\al{\alpha}
\def\de{\delta}\def\De{\Delta}\def\ep{\epsilon}
\def\te{\theta}\def\la{\lambda}
\def\La{\Lambda}\def\si{\sigma}

\def\vec#1{{\bf #1}}

\def\H{{_{H}}}\def\In{{_{in}}}
\newcommand{\mlab}[1]{\label{#1}}
\begin{document}%

\title{Quantum fields with topological defects}

\vspace{1.0cm}

\author{Massimo Blasone${}^{\sharp}$, Petr Jizba${}^{\flat}$ and Giuseppe Vitiello${}^{\sharp}$}

\vspace{1.0cm}

\address{${}^{\sharp}$ Dipartimento di Fisica ``E.R.Caianiello'', Universit\`a di Salerno,
\\ INFN and INFM, Salerno 84100 Italy
\\ [1mm] ${}^{\flat}$ Institute for Theoretical Physics, University of Tsukuba,
Ibaraki 305-8571, Japan}


\maketitle

\begin{abstract}
Domain walls, strings and monopoles are extended objects, or
defects, of quantum origin with topologically non--trivial
properties and macroscopic behavior. They are described in Quantum
Field Theory in terms of inhomogeneous condensates. We review the
related formalism in the framework  of the spontaneous breakdown
of symmetry.
\end{abstract}

\section{Introduction}

The ordered patterns we observe in condensed matter  and in high
energy physics are created by the quantum dynamics. Macroscopic
systems exhibiting some kind of ordering, such as superconductors,
ferromagnets, crystals, are described by the underlying quantum
dynamics. Even the large scale structures in the Universe, as well
as the ordering in the biological systems appear to be the
manifestation of the microscopic dynamics ruling the elementary
components of these systems. Thus we talk  of {\it macroscopic
quantum systems}: these are quantum systems in the sense that,
although they behave classically, nevertheless some of their
macroscopic features cannot be understood without recourse to
quantum theory.

The question then arises of how the quantum dynamics generates the
observed macroscopic properties. In other words, how it happens
that the {\it macroscopic} scale characterizing those systems is
dynamically generated out of the {\it microscopic} scale of the
quantum elementary components\cite{Um1}.

Moreover, we also observe a variety of phenomena where quantum
particles coexist and interact with {\it extended macroscopic
objects} which show a classical behavior, e.g. vortices in
superconductors and superfluids, magnetic domains in ferromagnets,
dislocations and other {\it topological defects} (grain
boundaries, point defects, etc.) in crystals, and so on.

We are thus faced also with the question of the quantum origin of
topological defects and of their interaction with
quanta\cite{Um1}: This is a crucial issue for the understanding of
symmetry breaking phase transitions and structure formation in a
wide range of systems from condensed matter to cosmology
\cite{kib,zu,vol}.

Here, we will review how the generation of ordered structures and
of extended objects is explained in Quantum Field Theory (QFT). We
follow refs.\cite{Um1} in our presentation. We will
consider systems in which spontaneous symmetry breaking (SSB)
occurs and show that topological defects originate by
inhomogeneous (localized) condensation of quanta.
The  approach followed here is
alternative to the usual one \cite{Raj}, in which one starts from
the classical soliton solutions and then ``quantizes'' them,
as well as to the QFT method based on dual (disorder) fields \cite{Klei}.

In Section 2 we first introduce some general features of QFT
useful for our discussion and we then treat some aspects of SSB
and the rearrangement of symmetry. In Section 3 we discuss the
boson transformation theorem and the topological singularities of
the boson condensate. Section 4 contains as an example a model
with $U(1)$ gauge invariance in which SSB, rearrangement of
symmetry and topological defects are present \cite{MPUV75}.
There we show how macroscopic fields and currents are obtained
from the microscopic quantum dynamics. The Nielsen-Olesen vortex
solution is explicitly obtained as an example. Section 5 is
devoted to conclusions.

\section{Symmetry and order in QFT: a dynamical problem}

QFT deals with systems with infinitely many degrees of freedom.
The  fields used for their description are {\it operator fields}
whose mathematical significance is fully specified only when the
state space where they operate is  also assigned.  This is the
space of the states, or physical phase, of the system under given
boundary conditions. A change in the boundary conditions may
result in the transition of the system from one phase to another
one. For example, a change of the temperature from above to below
the critical temperature may induce the transition from the normal
to the superconducting phase in a metal. The identification of the
state space where the field operators have to be realized is thus
a physically non trivial problem in QFT. In this respect, the QFT
structure is drastically different from the one of Quantum
Mechanics (QM). The reason is the following.

The von Neumann theorem in QM \cite{vn} states that for systems
with a finite number of degrees of freedom all the irreducible
representations of the canonical commutation relations are
unitarily equivalent. Therefore in QM the physical system can only
live in one single physical phase: unitary equivalence means
indeed physical equivalence and thus there is no room (no
representations) for physically different phases. Such a situation
drastically changes in QFT where systems with infinitely many
degrees of freedom are treated. In such a case the von Neumann
theorem does not hold and infinitely many unitarily inequivalent
representations of the canonical commutation relations do in fact
exist\cite{bog,brat}. It is such a richness of QFT which allows
the description of different physical phases.

\subsection{QFT as a two-level theory}

In the  perturbative approach, any quantum experiment or
observation can  be schematized as a scattering process where one
prepares a set of free (non-interacting) particles (incoming
particles or in-fields) which are then made to collide at some
later time in some space region (space-time region of
interaction). The products of the collision are expected to emerge
out of the interaction region as free particles (outgoing
particles or out-fields). Correspondingly, one has the in-field
and the out-field state space. The interaction region is where the
dynamics operates: given the in-fields and the in-states, the
dynamics determines the out-fields and the out-states.

The incoming particles and the outgoing ones (also called
quasi-particles in solid state physics)  are well
distinguishable and localizable particles only far away from the
interaction region, at a time much before ($t=- \infty$) and much
after ($t=+ \infty$) the interaction time: in- and out-fields are
thus said to be asymptotic fields, and for them the interaction
forces are assumed not to operate (switched off).

The only regions accessible to observations are those  far away
(in space and in time) from the interaction region, i.e. the
asymptotic regions (the in- and out-regions). It is so since, at
the quantum level, observations performed in the interaction
region or vacuum fluctuations there occurring may drastically
interfere with the interacting objects thus changing their nature.
Besides the asymptotic
fields, one then also introduces dynamical or Heisenberg fields, i.e.
the fields in terms of which the dynamics is given.
Since the interaction region
is precluded from observation, we do not observe Heisenberg
fields. Observables are thus solely described in terms of
asymptotic fields.

Summing up, QFT is a ``two-level'' theory: one level is the
interaction  level where the dynamics is specified by
assigning the equations for the Heisenberg fields. The other level
is the physical level, the one of the asymptotic fields and of the
physical state space directly accessible to observations. The
equations for the physical fields are equations for free fields,
describing the observed incoming/outgoing particles.

To be specific, let the Heisenberg operator fields
be generically denoted by $\psi_\H(x)$ and the physical operator
fields by $\varphi_\In (x)$\footnote{For definiteness,
we choose to work with the in-fields, although the set of
out-fields would work equally well.}. They are both assumed to satisfy
equal-time canonical (anti-)commutation relations.

For shortness, we omit considerations on the renormalization
procedure, which are not essential for the conclusions we will
reach. The Heisenberg field equations and the free field equations
are written as
\bea \lab{bt1} \La(\pa) \, \psi_\H(x)& =& {\mathcal{J}}[\psi_\H](x) \\ [2mm]
\lab{bt2} \La(\pa) \, \varphi_\In(x)& = & 0 \, ,  \eea
where  $\La(\pa)$ is a differential operator, $x\equiv
(t,\vec{x})$ and ${\mathcal{J}}$ is   some functional  of  the $\psi_\H$ fields,
describing the interaction.

Eq.(\ref{bt1}) can be formally recast in the following
integral form (Yang--Feldman equation):
\bea \lab{bt3}
\psi_\H(x)\, =\, \varphi_\In(x)\, + \, \La^{-1}(\pa)\ast \,{\mathcal{J}}[\psi_\H](x) \,
,  \eea
\noindent where  $\ast$  denotes convolution. The symbol
$\La^{-1}(\pa)$ denotes formally  the   Green function for
$\varphi_\In(x)$.  The precise form of Green's function is
specified by the boundary conditions.  Eq.(\ref{bt3}) can  be
solved by iteration, thus giving an expression for the Heisenberg
fields $\psi_\H(x)$ in terms of powers of the $\varphi_\In(x)$
fields; this is the Haag expansion in the LSZ
formalism\cite{bog,itz} (or ``dynamical map'' in the language of
refs.\cite{Um1}), which might be formally written
as\footnote{A (formal) closed form for the dynamical map is
obtained in the closed time path (CTP) formalism \cite{BJ02}.
Then the Haag expansion
(\ref{bt4}) is  directly applicable to
 both equilibrium and non--equilibrium situations.}
\bea  \lab{bt4} \psi_\H(x)\, =\, F \lf[x; \varphi_\In \ri] \, .  \eea

We stress that the equality in the dynamical map (\ref{bt4}) is a
``weak'' equality, which means that it must be understood as an
equality among matrix elements computed in the Hilbert space of the physical
particles.

We observe that mathematical consistency in the above procedure
requires that the set of $\varphi_\In$ fields must be an
irreducible set; however, it may happen that not all the elements
of the set are known since the beginning. For example there might
be composite (bound states) fields or even elementary quanta whose
existence is ignored in a first recognition. Then the computation
of the matrix elements in physical states will lead to the
detection of unexpected poles in the Green's functions, which
signal the existence of the ignored quanta. One thus introduces
the fields corresponding to these quanta and repeats the
computation. This way of proceeding is called the self-consistent
method\cite{Um1}. We remark that it is not necessary to have a
one-to-one correspondence between the sets $\{\psi_\H^j\}$ and
$\{\varphi_\In^i\}$,  as it happens whenever
the set $\{\varphi_\In^i\}$ includes composite particles.

\subsection{The dynamical rearrangement of symmetry}

As already mentioned, in QFT the Fock space for the physical
states is not unique since one may have several physical phases,
e.g. for a metal the normal phase and the superconducting phase,
and so on. Fock spaces describing different phases are unitarily
inequivalent spaces and correspondingly we have different
expectation values for certain observables and even different
irreducible sets of physical quanta. Thus, finding the dynamical map involves
singling out the Fock space where the dynamics has to be realized.

Let us now suppose that the Heisenberg field equations are invariant under some group
$G$ of transformations of $\psi_\H$:
\be\mlab{rs1} \psi_\H(x) \rar \psi_\H'(x) = g\lf[ \psi_\H(x) \ri]~, \ee
with $g \in G$.
The symmetry is spontaneously broken when the vacuum
state in the Fock space ${\cal H}$ is not invariant under the
group $G$ but only under one of its subgroups\cite{Um1,bog,itz}.

On the other hand, Eq.(\ref{bt4}) implies that when $\psi_\H$ is
transformed as in (\ref{rs1}), then
\be\mlab{rs3} \varphi_\In(x) \rar \varphi_\In'(x) = g'\lf[ \varphi_\In(x)
\ri]~, \ee
with $g'$ belonging to some group of transformations $G'$ and such that
\be\mlab{rs4} g\lf[ \psi_\H(x)\ri] = F\lf[ g'\lf[\varphi_\In(x)\ri]
\ri]~. \ee

When symmetry is spontaneously broken it is $G' \neq G$, with $G'$
the group contraction of $G$\cite{dv};
when symmetry is not broken $G' = G$.

Since $G$ is the invariance group of the dynamics, Eq.(\ref{bt4})
requires that $G'$ is the group under which free fields equations
are invariant, i.e. also $\varphi_\In'$ is a solution of
(\ref{bt2}). Since Eq.(\ref{bt4}) is a weak equality, $G'$ depends
on the choice of the Fock space ${\cal H}$ among the physically
realizable unitarily inequivalent state spaces. Thus we see that
the (same) original invariance of the dynamics may manifest itself
in different symmetry groups for the $\varphi_\In$ fields
according to different choices of the physical state space. Since
this process is constrained by the dynamical equations
(\ref{bt1}), it is called the {\it dynamical rearrangement of
symmetry}\cite{Um1}.

In conclusion, different ordering patterns appear to be different
{\it manifestations} of the same basic dynamical invariance. The
discovery of the process of the {\it dynamical rearrangement of
symmetry} leads to a unified understanding of the dynamical
generation of many observable ordered patterns.  This is the
phenomenon of the {\it dynamical generation of order}. The
contraction of the symmetry group is the mathematical
structure controlling the dynamical rearrangement of the
symmetry\cite{dv}. For a qualitative presentation see
Ref.\cite{Double}.

One can now ask which ones are the carriers of the ordering
information among the system elementary constituents and  how the
long range correlations and the coherence observed in ordered
patterns are generated and sustained. The answer is in the fact
that  SSB implies the appearance of boson particles\cite{go,nj},
the so called Nambu-Goldstone (NG) modes or quanta. They manifest
as long range correlations and thus they are responsible of the
above mentioned change of scale, from microscopic to macroscopic.
The coherent boson condensation of NG modes turns out to be the
mechanism by which order is generated, as we will see in an
explicit example in Section 4.

\section{The ``boson transformation'' method}

We now discuss the quantum origin of extended objects (defects)
and show how they naturally emerge as macroscopic objects
(inhomogeneous condensates) from the quantum dynamics. At zero temperature, the
classical soliton solutions
are then recovered in the Born
approximation. This approach is known as the ``boson
transformation'' method \cite{Um1}.

\subsection{The boson transformation theorem}

Let us consider, for simplicity, the case of a dynamical model
involving one scalar field $\psi_\H$ and one asymptotic field
$\varphi_\In$ satisfying Eqs.(\ref{bt1}) and  (\ref{bt2}),
respectively.

As already remarked, the dynamical map is valid only in a weak
sense, i.e. as a relation among matrix elements. This implies that
Eq.(\ref{bt4}) is not unique, since different sets of asymptotic
fields and the corresponding Hilbert spaces can be used in its
construction. Let us indeed consider a c--number function $f(x)$,
satisfying the $\varphi_\In$ equations of motion (\ref{bt2}):
\bea \lab{bt5}
\La(\pa) \, f(x)\, = \, 0 \, .
\eea
The {\em boson transformation theorem} \cite{Um1} states that the
field
\bea \lab{bt7} \psi_\H^f(x)\, =\, F \lf[x; \varphi_\In +f\ri]\, .
\eea
is also a solution of the Heisenberg equation (\ref{bt1}). The
corresponding Yang--Feldman equation takes the form
\bea \lab{bt6} \psi_\H^f(x)\, =\,  \varphi_\In(x)\,  + \, f(x)\, + \,
\La^{-1}(\pa) \ast {\mathcal{J}}[\psi_\H^f](x)\, .
\eea
 The difference between the two solutions $\psi_\H$
and $\psi_\H^f$ is only in the boundary conditions. An important
point is that the expansion Eq.(\ref{bt7}) is obtained from that
in Eq.(\ref{bt4}), by the space--time dependent translation

\bea \lab{bt7b} \varphi_\In(x) \,\rar \,\varphi_\In(x) \,+\,f(x)\,
. \eea
The essence of the  boson transformation theorem is that the
dynamics embodied in Eq.(\ref{bt1}), contains an internal freedom,
represented by the possible choices of the function $f(x)$,
satisfying the  free field  equation (\ref{bt5}).

We also observe that the transformation (\ref{bt7b}) is a
canonical transformation since it leaves invariant the canonical
form of commutation relations.


Let $|0\ran $ denote the vacuum for the free field $\varphi_\In$.
The vacuum expectation value   of  Eq.(\ref{bt6}) gives:
\bea \lab{bt8}
\phi^f(x)&\equiv & \,\lan 0| \psi_\H^f(x) |0\ran \; = \;
f (x)\,+ \,\lan 0|\lf[ \La^{-1}(\pa) \ast {\mathcal{J}}
[\psi_\H^f](x) \ri]|0\ran \, .  \eea
The c--number field $\phi^f(x)$ is the  order  parameter. We
remark that it is fully determined by the quantum dynamics. In the
classical or Born approximation, which consists in taking $\,\lan
0|{\mathcal{J}}[\psi_\H^f]|0\ran= {\mathcal{J}}[\phi^f]$, i.e.
neglecting all the contractions of the physical fields, we define
$\phi^f_{cl}(x)\equiv \lim_{\hbar\rar 0}\phi^f(x)$. In this limit
we have
\bea \lab{bt9} \La(\pa) \, \phi^f_{cl}(x)\, =\,
{\mathcal{J}}[\phi^f_{cl}](x) \, ,
\eea
i.e. $\phi^f_{cl}(x)$ provides  the solution of the
classical Euler--Lagrange equation.

Beyond the classical level, in general, the form of this equation
changes. The Yang--Feldman equation (\ref{bt6}) gives not only
the equations for the order parameter Eq.(\ref{bt9}), but
also, at higher orders in $\hbar$, the dynamics  of the physical
quanta in the potential generated by the ``macroscopic object''
$\phi^f(x)$ \cite{Um1}.

One can show\cite{Um1}, that the class of solutions of
Eq.(\ref{bt5}) which lead to topologically non--trivial (i.e.
carrying a non--zero topological charge) solutions of
Eq.(\ref{bt9}), are those which have some sort of singularity with
respect to Fourier transform. These can be either  {\em divergent
singularities} or {\em topological singularities}.  The first are
associated to a divergence of $f(x)$ for $|\vec{x}|=\infty$, at
least  in some direction.  Topological singularities are instead
present when $f(x)$ is not single-valued, i.e. it is path
dependent. In both cases, the macroscopic  object described by the
order parameter, carries a non--zero topological charge.

\subsection{Topological singularities and massless bosons}

An important result is that the boson transformation functions
carrying topological singularities are only allowed for massless
bosons\cite{Um1}.

Consider a  generic boson field $\chi_\In$ satisfying the equation
\be\mlab{kg} (\pa^2 + m^2)\chi_\In(x) = 0~, \ee
and suppose that the function $f(x)$ for the boson transformation $\chi_\In(x) \rar
\chi_\In(x) + f(x)$ carries a topological singularity. It is
then  not single-valued and thus path-dependent:
\be
\\ \mlab{ts1}
G^{+}_{\mu\nu}(x) \equiv [\pa_\mu,\pa_\nu]\,f(x) \neq 0~, \qquad
{\rm for \;certain }\; \mu\, , \,  \nu \, , \, x  ~. \ee
On the other hand, $\pa_\mu \, f(x)$, which is related with
observables, is single-valued, i.e. $[\pa_\rho,\pa_\nu]\,\pa_\mu
f(x)\,=\,0$. Recall that $f(x)$ is solution of the $\chi_\In$
equation:
\be\mlab{ts3} (\pa^2 + m^2)f(x) = 0~. \ee
{}From the definition of $G^{+}_{\mu\nu}(x)$ and the regularity of
$\pa_\mu f(x)$ it follows, by computing $\pa^{\mu}
G^{+}_{\mu\nu}(x)$, that
\be\mlab{ts5} \pa_\mu f(x) \, =\, \frac{1}{\pa^2 \, + \, m^2}
\pa^{\la} \,G^{+}_{\la\mu}(x) ~. \ee
This equation and the antisymmetric nature of $G^{+}_{\mu\nu}(x)$
then lead to $\pa^2 f(x) \, =\,0$, which in turn implies $m=0$.
Thus we conclude that (\ref{ts1}) is only compatible with massless
equation for $\chi_\In$.

The topological charge is defined as
\be\mlab{ts8} N_T=\int_C \,dl^\mu\, \pa_\mu \, f = \int_S\,
dS_{\mu}\ep^{\mu\nu\si}\, \pa_\nu \pa_\si \, f\, =\,
\frac{1}{2}\int_S\, dS^{\mu\nu}\, G^{+}_{\mu\nu} ~. \ee
Here $C$ is a contour enclosing the singularity and $S$ a surface
with $C$ as boundary. $N_T$ does not depend  on the path $C$
provided this does not cross the singularity. The dual tensor
$G^{\mu\nu}(x)$ is
\be\mlab{ts9} G^{\mu\nu}(x)\equiv - \frac{1}{2}\,
\ep^{\mu\nu\la\rho}G^{+}_{\la\rho}(x) \ee
and satisfies the continuity equation:
\be\mlab{ts10} \pa_\mu\,G^{\mu\nu}(x)\,=\,0 \qquad \Lrar \qquad
\pa_\mu\,G^{+}_{\la\rho}(x)\, +\,\pa_\rho\,G^{+}_{\mu\la}(x)\, +\,
\pa_\la\,G^{+}_{\rho\mu}(x)\,=\,0   ~. \ee
Eq.(\ref{ts10}) completely characterizes the topological singularity
\cite{Um1}.

\section{An example: The Anderson-Higgs-Kibble mechanism and the vortex solution}

We consider a model of a complex scalar field $\phi(x)$
interacting with a gauge field $A_\mu(x)$\cite{and,hig,ki}. The
lagrangian density ${\cal L}[\phi(x), \phi^*(x), A_\mu(x)]$ is
invariant under the global and the local $U(1)$
gauge transformations\footnote{We do not assume a particular form for the Lagrangian
density, so the following results are quite general.}:
\be\mlab{lp1}
\phi(x) \rar e^{i \te} \phi(x) \qquad, \qquad \qquad
A_\mu(x) \rar A_\mu(x)~,
\ee
\be\mlab{lp2}
\phi(x) \rar e^{i e_0 \la(x)} \phi(x) \qquad, \qquad \qquad
A_\mu(x) \rar A_\mu(x) \, + \, \pa_\mu \la(x)~,
\ee
respectively, where $\la(x)\rar 0$ for $|x_0|\rar \infty$ and/or
$|{\bf x}|\rar \infty$ and $e_0$ is the coupling constant. We work
in the Lorentz gauge $\pa_\mu A^\mu(x)\,=\,0$.
The generating functional, including the gauge
constraint, is \cite{MPUV75}
\bea \mlab{lp55} && {\cal Z}[J,K]\ = \ \frac{1}{\cal N}\int [d
A_\mu] [d\phi] [d\phi^*] [d B]\, \exp\lf[i \
{\mathcal{S}}[A_\mu,B, \phi]\ri]\, ,
\\  [2mm]\non
&& {\cal S}\ = \ \int d^{4} x \, \left[ {\cal L}(x) + B(x) \pa^\mu
A_\mu(x) + K^*(x) \phi(x) + K(x) \phi^*(x) +J^\mu(x) A_\mu(x)+ i {\ep}
|\phi(x) -v|^2 \right] \, ,
\\  [2mm] \non
&&{\cal N} \ = \ \int [d A_\mu] [d\phi] [d\phi^*] [d B]\, \exp\lf[i\int
d^{4} x\lf({\cal L}(x) + i {\ep} |\phi(x) -v|^2 \ri)\ri]\, . \eea
$B(x)$ is an auxiliary field which implements the
gauge fixing condition\cite{MPUV75,OJ90}.
Notice the $\ep-$term where $v$ is a complex number. Its r\^ole is
to specify the condition of symmetry breaking under which we want
to compute the functional integral and it may be
given the physical meaning of a small external field triggering
the symmetry breaking \cite{MPUV75}. The limit $\ep \rar 0$ must be made at the
end of the computations. We will use the notation
\be\mlab{gp2} \lan F[\phi]\ran_{\ep,J,K} \equiv \frac{1}{\cal N}\int [d
A_\mu] [d\phi] [d\phi^*] [d B]\, F[\phi]\, \exp\lf[i  \
{\mathcal{S}}[A_\mu,B, \phi] \ri]\, , \ee
with $\lan F[\phi]\ran_{\ep} \equiv  \lan F[\phi]\ran_{\ep, J=K=0}$ and
$\lan F[\phi]\ran \equiv  \lim_{\ep \rar 0} \, \lan F[\phi]\ran_{\ep}$.

The fields $\phi$, $A_\mu$ and $B$ appearing in the generating
functional are c-number fields. In the following the Heisenberg
operator fields corresponding to them will be denoted by
$\phi_\H$, $A_{\H \mu}$ and $B_\H$, respectively. Thus the
spontaneous symmetry breaking condition is expressed by $\lan
0|\phi_\H(x)|0\ran \equiv {\tilde v} \neq 0$, with ${\tilde v}$
constant.

Since in  the functional integral formalism the functional average
of a given c-number field gives the vacuum expectation value of
the corresponding operator field, e.g. $\lan F[\phi]\ran \equiv
\lan 0|F[\phi_\H]|0\ran$,  we have $\lim_{\ep \rar 0} \, \lan
\phi(x) \ran_{\ep} \equiv \lan 0|\phi_\H(x)|0\ran = {\tilde v}$.

Let us introduce the following decompositions:
$\phi(x)=\frac{1}{\sqrt{2}}\lf[\psi(x) + i \chi(x)\ri]$,
 $K(x)=\frac{1}{\sqrt{2}}\lf[K_1(x) + i
K_2(x)\ri]$ and
$\rho(x) \equiv \psi(x) - \lan\psi(x)\ran_\ep$. Note that
$\lan\chi(x)\ran_\ep=0$ because of the invariance under $\chi\rar
- \chi$.

\subsection{The Goldstone theorem}

Since the functional integral (\ref{lp55}) is invariant under the
global transformation (\ref{lp1}), we have that $\pa {\cal Z}[J,K]/ \pa\te =0$
and subsequent derivatives with
respect to $K_{1}$ and $K_{2}$ lead to
\be\mlab{gp6} \lan \psi(x)\ran_\ep = \sqrt{2}\,\ep v \int d^4y
\,\lan \chi(x)\chi(y)\ran_{\ep} = \sqrt{2}\,\ep v\,
\De_\chi(\ep,0)~. \ee
In momentum space the propagator for the field $\chi$ has the
general form
\be\mlab{gp8}
\De_\chi(0,p) = \lim_{\ep \rar 0}\lf[ \frac{Z_\chi}{p^2-m_\chi^2+i\ep
a_\chi} + {\rm (continuum \;contributions)} \ri]\, .
\ee
Here $Z_\chi$ and $a_\chi$ are renormalization constants.
The integration in Eq.(\ref{gp6}) picks up the pole contribution at
$p^2=0$, and leads to
\bea \mlab{gp9}
{\ti v}= \sqrt{2}\frac{Z_\chi}{a_\chi} v \; \Lrar \; m_\chi = 0  \qquad; \qquad
{\ti v}= 0 \; \Lrar\; m_\chi \neq 0   ~.
\eea
The Goldstone theorem \cite{go} is thus proved:
if the symmetry is spontaneously broken (${\ti v} \neq 0$), a
massless mode must exist, whose field is $\chi(x)$, i.e. the NG boson
mode. Since it is massless it manifests as a long range
correlation mode. (Notice that in the present case of a complex
scalar field model the NG mode is an elementary field . In other
models it may appear as a bound state, e.g. the magnon in
(anti-)ferromagnets). Note that
\be\mlab{gp11} \frac{\pa}{\pa v}\lan \psi(x) \ran_\ep = \sqrt{2}\,
\ep \int d^4 y \lan \rho(x)\rho(y)\ran_\ep~, \ee
and because $m_\rho\neq 0$, the r.h.s. of this equation vanishes
in the limit $\ep\rar 0$; therefore ${\ti v}$ is independent of $|v|$,
although the phase of  $|v|$ determines the one of ${\ti v}$ (from
Eq.(\ref{gp6})): as in ferromagnets, once an
external magnetic field is
switched on, the system is magnetized independently of the
strength of the external field.

\subsection{The dynamical map and the field equations}

Observing that the change of variables
(\ref{lp1}) (and/or (\ref{lp2}) ) does not affect the  generating functional,
we may obtain the Ward-Takahashi identities. Also, using
$B(x) \rar B(x) + \la(x)$ in (\ref{lp55}) gives
$\lan \pa^\mu A_\mu(x)\ran_{\ep,J,K} = 0$.
One then finds the following two-point function pole
structures\cite{MPUV75}:
\bea\mlab{lp16} &&\lan  B(x) \chi(y)\ran\, =\, \lim_{\ep\rar 0}
\lf\{ \frac{-i}{(2\pi)^4} \int d^4p\, e^{-i p (x-y)}\frac{e_0 {\ti
v}}{p^2 +i \ep a_\chi} \ri\}\, ,
\\ [2mm]
\mlab{lp17}&&\lan  B(x) A^\mu(y)\ran\, =\, \pa^\mu_x
\frac{i}{(2\pi)^4} \int d^4p\, e^{-i p (x-y)}\frac{1}{p^2 }\, ,
\\ [2mm]
\mlab{lp18} &&\lan  B(x) B(y)\ran\, =\, \lim_{\ep\rar 0} \lf\{
\frac{-i}{(2\pi)^4} \int d^4p\, e^{-i p (x-y)}\frac{(e_0 {\ti
v})^2}{Z_\chi} \lf[\frac{1}{p^2 +i \ep a_\chi} - \frac{1}{p^2 }
\ri]\ri\}  ~.
\eea
The absence of branch cut singularities in propagators
(\ref{lp16})--(\ref{lp18}) suggests that $B(x)$ obeys a free field
equation. In addition, Eq.(\ref{lp18}) indicates that the model
contains a massless negative norm state (ghost) besides the NG
massless mode $\chi$. Moreover, it can be shown \cite{MPUV75} that
a massive vector field $U_\In^{\mu}$ also exists in the theory. Note
that because of the invariance $(\chi,A_\mu,B) \rar
(-\chi,-A_\mu,-B)$, all the other two-point functions must vanish.

The dynamical maps expressing the Heisenberg operator fields in
terms of the asymptotic operator fields, are found to be
\cite{MPUV75}:
\bea\mlab{lp57a} &&\phi_\H(x)= :\exp\lf\{i
\frac{Z_\chi^{\frac{1}{2}}} { {\ti v}}\chi_\In(x) \ri\} \lf[{\ti
v} + Z_\rho^{\frac{1}{2}} \rho_\In(x) + {\mathcal{F}}[\rho_\In,
U^\mu_\In, \pa(\chi_\In - b_\In)] \ri]:\, ,
\\ [2mm] \mlab{lp57b}
&& A^{\mu}_{\H}(x)= {Z_3^{\frac{1}{2}}} U^{\mu}_\In(x) +
{\frac{Z_\chi^{\frac{1}{2}}}{e_0{\ti v}}}\  \pa^\mu b_\In(x) +
:{\mathcal{F}}^{\mu}[\rho_\In, U^\mu_\In, \pa(\chi_\In - b_\In)] :\, ,
\\ [2mm]\mlab{lp43}
&& B_\H(x)= \frac{e_0 {\ti v}}{Z_\chi^{\frac{1}{2}}}\ [b_\In(x) -
\chi_\In(x)] + c \, ,
 \eea
where $:...:$ denotes the normal ordering and the functionals
${\mathcal{F}}$ and ${\mathcal{F}}^\mu$ are to be determined
within a particular model. In Eqs.(\ref{lp57a})--(\ref{lp43}),
$\chi_\In$ denotes the NG mode, $b_\In$ the ghost mode,
$U^{\mu}_\In$ the massive vector field and $\rho_\In$ the massive
matter field. In Eq.(\ref{lp43}) $c$ is a c--number constant,
whose value is irrelevant since only derivatives of $B$ appear in
the field equations (see below). $Z_3$ represents the wave
function renormalization for $U^{\mu}_\In$. The corresponding
field equations are
\bea\mlab{lp24} && \pa^2 \chi_\In(x)\,=\,0\, ,~~~ \pa^2
b_\In(x)\,=\,0\, , ~~~ (\pa^2 \, + \, m_\rho^2)\rho_\In(x) \,
=\,0\, ,
\\ [2mm] \mlab{lp29}
&& ( \pa^2 \, + \, {m_V}^{2}) U_\In^\mu(x) \, =\, 0 \, , ~~~
\pa_\mu U_\In^{\mu}(x) \, =\, 0\, . \eea
with ${m_V}^{2} = \frac{Z_{3}}{Z_\chi} \ (e_0{\ti v})^{2}$.
The field equations for $B_{\H}$ and $A_{\H \mu}$
read\cite{MPUV75}
\be\mlab{lp37} \pa^2 B_\H(x)\, =\,0\, ,~~~ - \pa^2 A_{\H \mu}(x)
\, =\, j_{\H \mu}(x) \, -\, \pa_\mu B_\H(x) ~, \ee
with $j_{\H \mu}(x)= \de{\cal L}(x)/\de A^{\mu}_\H(x)$.
One may then require that the current $j_{\H \mu}$ is the only source of
the gauge field $A_{\H \mu}$ in any observable process. This amounts to
impose the condition: $_p\lan b|\pa_\mu B_\H(x)|a\ran_p\,
= \,0$, i.e.
\be\mlab{lp45}
(- \pa^2) \,_p\lan b| A^{0}_{\H \mu}(x) |a \ran_p \, =
\,_p\lan b| j_{\H \mu}(x) |a\ran_p  ~,
\ee
where $|a\ran_p $ and   $|b\ran_p $ denote two generic {\em
physical} states and $A^{0\mu}_{\H}(x) \equiv A^{\mu}_{\H}(x) - {
e_0{\ti v}}:\pa^\mu b_\In(x):$. Eq.(\ref{lp45}) are the classical
Maxwell equations. The condition $_p\lan b|\pa_\mu
B_\H(x)|a\ran_p\, = \,0$ leads to the Gupta--Bleuler--like
condition
\be\mlab{lp49}
[\chi_\In^{(-)}(x)  \, - \,  b_\In^{(-)}(x)]|a\ran_p\, = \,0   ~,
\ee
where $\chi_\In^{(-)}$ and $b_\In^{(-)}$ are the
positive--frequency parts of the corresponding fields. Thus we see
that $\chi_\In$ and $b_\In$ cannot participate in any observable
reaction. This is confirmed by
 the fact that they are present in the $S$ matrix in the combination $(\chi_\In -
b_\In)$ \cite{MPUV75}. It is to be remarked however that the NG
boson does not disappear from the theory: we shall see
below that there are situations in which  the NG fields do have
observable effects.

\subsection{The dynamical rearrangement of symmetry and the
classical fields and currents}

{}From Eqs.(\ref{lp57a})-(\ref{lp57b}) we see that the local gauge
transformations of the Heisenberg fields
\be\mlab{lp50}
\phi_\H(x) \rar e^{i e_0 \la(x)} \phi_\H(x)\,,\quad
A_\H^\mu(x) \rar A_\H^\mu(x) \, + \, \pa^\mu \la(x)\,,\quad
 B_\H(x) \rar B_\H(x),
\ee
with $\pa^2\la(x) =0$, are induced by the in-field transformations
\bea\non
&&\chi_\In(x)  \rar   \chi_\In(x) \, + \, \frac{e_0 {\ti
v}}{Z_\chi^{\frac{1}{2}}} \la(x)\,,\quad
b_\In(x)  \rar   b_\In(x) \, + \, \frac{e_0 {\ti
v}}{Z_\chi^{\frac{1}{2}}} \la(x)\,,
\\ [2mm]\mlab{lp51a}
&&\rho_\In(x)  \rar   \rho_\In(x)\,,\quad
U^\mu_\In(x) \rar  U^\mu_\In(x)  ~.
\eea
On the other hand, the global phase transformation $\phi_\H(x) \rar e^{i  \te} \phi_\H(x)$
is induced by
\bea\non
&&\chi_\In(x)  \rar  \chi_\In(x) \, + \, \frac{{\ti
v}}{Z_\chi^{\frac{1}{2}}} \te f(x) \quad, \quad b_\In(x)  \rar   b_\In(x) ~,
\\ [2mm] \mlab{lp53b}
&&
\rho_\In(x)  \rar   \rho_\In(x) \quad, \quad
U^\mu_\In(x)  \rar   U^\mu_\In(x)  ~,
\eea
with $\pa^2 f(x) =0$ and the limit $f(x)\rar 1$  to be performed
at the end of computations. Note that under the above
transformations the in-field equations and the $S$ matrix  are
invariant and that $B_\H$ is changed by an irrelevant c-number (in
the limit $f \rar 1$ ).

Consider now the boson transformation
$\chi_\In(x)  \rar  \chi_\In(x) \, + \, \al (x)$:
In local gauge theories the boson transformation must be
compatible with the Heisenberg field equations but also with the
physical state condition (\ref{lp49}). Under the boson
transformation with $\al (x) =  {\ti
v} Z_\chi^{-\frac{1}{2}} \te f(x)$ and $\pa^2 f(x) =0$, $B_\H$
changes as
\be\mlab{vs9}
B_\H(x) \rar B_\H(x) - \frac{e_0 {\ti v}^2}{Z_\chi} f(x) ~,
\ee
Eq.(\ref{lp45}) is thus violated when the
Gupta-Bleuler-like condition is imposed. In order to
restore it, the shift in $B_\H$ must be compensated by means of
the transformation on $U^{\mu}_\In$:
\be\mlab{vs10}
U^{\mu}_\In(x) \rar U^{\mu}_\In(x) +
{Z_{3}}^{-\frac{1}{2}} a^{\mu}(x) \qquad ,
\qquad \pa_\mu a^{\mu}(x)=0 ~,
\ee
with a convenient c-number function $a^{\mu}(x)$. The dynamical
maps of the various Heisenberg operators are not affected by
(\ref{vs10}) since they contain $U^{\mu}_\In$ and $B_\H$ in a
combination such that the changes of $B_\H$ and of $U^{\mu}_\In$
compensate each other provided
\be\mlab{vs20}
(\pa^2 + m_V^2) a_\mu(x) \, = \,\frac{m_V^2}{ e_0} \pa_\mu f(x)~.
\ee
Eq. (\ref{vs20}) thus obtained is the Maxwell equation for
the massive potential vector $a_{\mu}$\cite{MPUV75}.
The classical ground
state current $j^\mu$ turns out to be
\be\mlab{vs21}
j^\mu(x)\equiv \lan 0| j_\H^\mu(x) |0 \ran \, =\,
 m_V^2 \lf[ a^\mu(x) - \frac{1}{e_0} \pa^\mu f(x) \ri]~.
\ee
The term $ m_V^2  a^\mu(x)$ is the {\em Meissner current}, while
$ \frac{m_V^2}{e_0} \pa^\mu f(x)$ is the {\em boson current}.
 The key point here is that both the  macroscopic field and current are given in terms of the
boson condensation function $f(x)$.

Two  remarks are in order: First, note that the terms proportional
to $\pa^\mu f(x)$ are related to observable effects, e.g. the
boson current which acts as the source of the classical field.
Second, note that the macroscopic ground state effects do not
occur for regular $f(x)$ ($G^{+}_{\mu\nu}(x) = 0$). In fact, from
(\ref{vs20}) we obtain $a_{\mu}(x) = \frac{1}{e_{0}} \pa_{\mu}
f(x)$ for regular $f(x)$ which implies zero classical current
($j_{\mu} = 0$) and zero classical field ($F_{\mu\nu} = \pa_{\mu}
a_{\nu} -  \pa_{\nu} a_{\mu}$), since the Meissner and the boson
current cancel each other.

In conclusion, the vacuum current appears only when $f(x)$ has
topological singularities and these can be created only by
condensation of massless bosons, i.e. when SSB occurs. This
explains why topological defects appear in the process of phase
transitions, where NG modes are present and gradients in their
condensate densities are nonzero \cite{kib,zu}.

On the other hand, the appearance of space-time order parameter
is no guarantee that persistent ground state currents (and
fields) will exist: if $f$(x) is a regular function, the space-time
dependence of $\ti v$ can be gauged away by an appropriate gauge
transformation.

Since, as said, the boson transformation with
regular $f(x)$ does not affect observable quantities, the $S$
matrix  is actually given by
\be\mlab{lp56aa}
S\,=\,: S[\rho_\In, U^\mu_\In - \frac{1}{m_V} \pa(\chi_\In -
 b_\In)] : ~.
\ee
This is indeed independent of the boson transformation with
regular $f(x)$:
\be\mlab{lp56ab}
S\,\rar\,S' = : S[\rho_\In, U^\mu_\In - \frac{1}{m_V}
\pa(\chi_\In - b_\In)
+ Z^{-\frac{1}{2}}_{3} (a^{\mu} - \frac{1}{e_{0}} {\pa}^{\mu} f)]:
\ee
since  $a_{\mu}(x) = \frac{1}{e_{0}}
\pa_{\mu} f(x)$ for regular $f(x)$. However, $S'
\neq S$ for singular $f(x)$: $S'$ includes the interaction of the quanta
$U^\mu_\In$ and $\phi_\In$ with the classically behaving macroscopic defects \cite{Um1}.

\subsection{The vortex solution}

Below we consider the example of the Nielsen--Olesen vortex string
solution. We show which one is the boson function $f(x)$
controlling the non--homogeneous NG boson condensation in terms of
which the string solution is described. For shortness, we only
report the results of the computations. The detailed derivation as
well as the discussion of further examples can be found in Ref.
\cite{Um1}.

In the present $U(1)$ problem, the electromagnetic tensor and the
vacuum current are\cite{Um1,MPUV75}
\bea\mlab{vs27}
 F_{\mu \nu}(x)& =&\pa_\mu a_{\nu}(x) - \pa_\nu
a_{\mu}(x)\, =\, 2\pi \frac{m_V^2}{e_0}\int d^4 x' \, \De_c(x-x')
G^{+}_{\mu \nu}(x')  ~,
\\ [2mm]
\mlab{vs28}
j_\mu(x) & =&
- 2\pi \frac{m_V^2}{e_0}\int d^4 x' \, \De_c(x-x') \pa_{x'}^\nu
G^{+}_{\nu \mu}(x') ~, \eea
respectively, and satisfy $\pa^\mu F_{\mu \nu}(x)\, =\,-j_\nu(x)$.
In these equations
\be \De_c(x-x') = \frac{1}{(2 \pi)^4} \int d^4 p \, e^{-i p
(x-x')} \,\frac{1}{p^2 - m_V^2 + i\ep} ~.
\ee

The line singularity for the vortex (or string) solution can be
parameterized by a single line parameter $\si$ and by the time
parameter $\tau$. A {\it static vortex} solution is obtained by
setting $y_0(\tau,\si)=\tau$ and ${\bf y}(\tau,\si)={\bf y}(\si)$,
with $y$ denoting the line coordinate. $G^{+}_{\mu \nu}(x)$ is
non-zero only on the line at $y$ (we can consider more lines but
let us limit to only one line, for simplicity). Thus, we have:
\bea\non
&&G_{0i}(x) \, =\,\int\, d\si\,
 \frac{d y_i(\si)}{d \si}
 \, \de^{3}[{\bf x} - {\bf y}(\si)] \,, \qquad
G_{ij}(x) \, =\,0 ~,
\\ [2mm]\mlab{vs31b}
&&G_{ij}^{+}(x) \, =\,-\ep_{ijk}G_{0k}(x) \,, \qquad
G_{0i}^{+}(x) \, =\,0 ~. \eea
Eq.(\ref{vs27}) shows that these vortices are purely {\em magnetic}.
We obtain
\be\mlab{vs32b}
\pa_{0}f(x) \, =\, 0 ~,~~~
\pa_{i}f(x) \, =\,\frac{1}{(2\pi)^2}\int\, d\si\,
\ep_{ijk} \frac{d y_k(\si)}{d \si} \pa_j^x \int d^3p
\frac{e^{i {\bf p}\cdot({\bf x}-{\bf y}(\si))}}{{\bf p}^2} ~,
\ee
i.e., by using the identity $(2\pi)^{-2}\int
d^3p \frac{e^{i {\bf p}\cdot{\bf x}}}{{\bf p}^2}= \frac{1}{2 |{\bf x}|}$,
\be
\mlab{vs33}
\nabla f(x) \, =\,-\frac{1}{2}\int\, d\si\,
\frac{d {\bf y}_k(\si)}{d \si} \wedge \nabla_x
\frac{1}{|{\bf x}-{\bf y}(\si)|} ~,
\ee
Note that $\nabla^2 f(x) =0$ is satisfied.

A straight infinitely long vortex is specified by $y_i(\si)= \si\,
\de_{i3}$ with $ -\infty < \si < \infty$. The only non vanishing
component of $G^{\mu\nu}(x)$ are $G^{03}(x)=G_{12}^{+}(x)\,=\,
\de(x_1)\de(x_2)$. Eq.(\ref{vs33}) gives\cite{Um1,MPUV75}
\beaa\non
\frac{\pa}{\pa x_1} f(x) &=&\frac{1}{2}\int d\si \frac{\pa}{\pa x_2}
[x_1^2 + x_2^2 + (x_3 -\si)^2]^{-\frac{1}{2}}=
- \frac{x_2}{x_1^2 + x_2^2} ~,
\\ [2mm]\mlab{vs36b}
 \frac{\pa}{\pa x_2} f(x) & =&\frac{x_1}{x_1^2 + x_2^2} ~,~~~
\frac{\pa}{\pa x_3} f(x)  = 0 ~,
\eeaa
and then
\be
\mlab{vs37}
f(x)\, =\, \tan^{-1}\lf(\frac{x_2}{x_1}\ri)\,=\,\te(x)   ~.
\ee
We have thus determined the   boson
transformation function corresponding
to a particular vortex solution.
The vector potential is
\beaa\non a_1(x) & =& -\frac{m_V^2}{2 e_{0}}\int d^4 x' \,
\De_c(x-x') \frac{x_2'}{x_1^{'2} + x_2^{'2}}\,,
\\ [2mm]\non
a_2(x) & =& \frac{m_V^2}{2 e_{0}}\int d^4 x' \, \De_c(x-x')
\frac{x_1'}{x_1^{'2} + x_2^{'2}}\,,
\\ [2mm]\mlab{vs38c}
a_3(x) & =&a_0(x) =0\,,
\eeaa
and the only non-vanishing component of $F_{\mu \nu}$:
\bea\mlab{vs39a} F_{12}(x)&=& - 2\pi \frac{m_V^2}{ e_{0}}\int d^4
x' \, \De_c(x-x') \de(x_1')\de(x_2') = \frac{m_V^2}{ e_{0}}\,
K_0\lf(m_V\sqrt{x_1^2 + x_2^2}\ri). \eea
Finally, the vacuum current Eq.(\ref{vs28}) is given by
\bea\non &&j_{1}(x)  = - \frac{m_V^3}{
e_{0}}\frac{x_2}{\sqrt{x_1^2 + x_2^2}} \,K_1\lf(m_V\sqrt{x_1^2 +
x_2^2}\ri)\,,
\\ [2mm]\non
&&j_{2}(x)=  \frac{m_V^3}{ e_{0}}\frac{x_1}{\sqrt{x_1^2 + x_2^2}}
\,K_1\lf(m_V\sqrt{x_1^2 + x_2^2}\ri)\,,
\\ [2mm]\mlab{vs40c}
&&j_{3}(x)=j_{0}(x)=0\,.
\eea
We observe that these results are the same of the Nielsen-Olesen vortex
solution \cite{NO}.
Notice that we did not  specify
the potential in our model but only  the invariance properties. Thus,
  the invariance properties of the dynamics determine the characteristics
of the topological solutions.  The vortex solution manifests the original  $U(1)$
symmetry through the cylindric angle $\te$ which is the parameter of the $U(1)$
representation in the coordinate space.

\section{Conclusions}

We have discussed how topological defects arise as inhomogeneous
condensates in Quantum Field Theory. Topological defects are shown
to have a genuine quantum nature. The approach reviewed here goes
under the name of ``boson transformation method'' and relies on
the existence of unitarily inequivalent representations of the
field algebra in QFT.

Describing quantum fields with topological defects amounts then to
properly choose the physical Fock space for representing the
Heisenberg field operators. Once the boundary conditions
corresponding to a particular soliton sector are found, then the
Heisenberg field operators embodied with such conditions contain
the full information about the defects, the quanta and their
mutual interaction. One can thus calculate Green's functions for
particles in the presence of  defects. The extension to finite
temperature is discussed in Refs.\cite{BJ02,MV90}.

As an example we have discussed a model with $U(1)$ gauge
invariance and SSB and we have obtained the Nielsen-Olesen vortex
solution\cite{NO} in terms of localized condensation of Goldstone
bosons. These thus appear to play a physical role, although, in
the presence of gauge fields, they do not show up in the physical
spectrum as excitation quanta. The function $f(x)$ controlling the
condensation of the NG bosons must be singular in order to produce
observable effects. Boson transformations with regular $f(x)$ only
amount to gauge transformations. For the treatment of topological
defects in non-abelian gauge theories, see Ref.\cite{MV90}.

Finally, when there are no NG modes, as in the case of the kink solution or
the sine-Gordon solution, the boson transformation function has to
carry divergence singularity at spatial infinity\cite{Um1,BJ02}.
In ref. \cite{MRV81} the boson transformation has been also
discussed in connection with the B\"aklund transformation at a
classical level and the confinement of the constituent quanta in
the coherent condensation domain.

For further reading on quantum fields with topological defects, see Ref.\cite{BJV}.

\smallskip

 We thank MIUR, INFN, INFM and the ESF network COSLAB for partial
 financial support.


\end{document}